\documentclass{article}
\usepackage{epsfig}
\usepackage{graphicx}
\usepackage{amsfonts}
\usepackage{euscript}
\usepackage{amsmath}
\usepackage{amssymb}
\begin{document}

\begin{center}

{\bf
\title "Stellar masses: the comparison of theoretical predictions and
measurement data}

\bigskip

\bigskip
\author "B.V.Vasiliev
\bigskip

\bigskip
\end{center}


\abstract {The Euler equation has been accepted as the basic
postulate of stellar physics long before the plasma physics was
developed. The existence of electrical interaction between particles
of interstellar plasma poses the question, how this interaction must
be accounted for. We argue that the right way is in formulation of a
new postulate. On the base of the new postulate, the theory of a hot
star interior is developed. Using this theory we obtain the
distribution of stars over their masses and mass-radius,
mass-temperature and mass-luminosity dependencies. All these
theoretical predictions are in a good agreement with the known
measurement data, which confirms the validity of the new postulate.}


\section{Introduction}
The astrophysics uses the Euler equation
\begin{equation} \gamma
\mathbf{g}-\nabla P=0\label{Eu}
\end{equation}
as its basic postulate at a stellar interior balance description. It
contends that a gravity force acting on substance inside star are
balanced by induced pressure gradient. During a long time this claim
seemed so obvious that nobody doubted its applicability to the
description of stellar substance equilibrium. Later on, as the
plasma physics has been developed, one could see that the particles
of stellar substance were possessing not only masses but also
electrical charges. The gravity can induce a redistribution of
electrical charges in plasma inside a star, i.e. produce an electric
polarization of plasma. The existence of this polarization must be
taken into account by formulating the equilibrium condition of
stellar interior. The force acting in a polarized substance is
proportional to its polarization $\mathfrak{P}$ and gradient
$\mathfrak{P}$ \cite{LL}. So the equilibrium equation taking into
account this force has the form:
\begin{equation}
\gamma \mathbf{g}+ 4\pi\mathfrak{P} \nabla\mathfrak{P}-\nabla P=0.
\end{equation}
The Eq.({\ref{Eu}}) describes the balance of non-polarized matter.
One could suppose that under certain conditions the equilibrium of
plasma in a star can be reached in another way, when a gravity force
is balanced by an induced electrical force. Therefore, instead of
the Eq.({\ref{Eu}) we postulate an alternative solution of this
problem:
\begin{equation}
\gamma \mathbf{g}+4\pi \mathfrak{P} \nabla\mathfrak{P}=0.\label{eP}
\end{equation}
The electrodynamic says that an electrical polarization can be
described as an appropriate bonded charge distribution
\begin{equation}
div\mathfrak{P}=\rho_{bond},
\end{equation}
and the equilibrium equation in this case can be formulated as
\begin{equation}
\gamma {\mathbf g} +\rho_{bond}~\mathbf{E} = 0\label{gen31}
\end{equation}
where the intensity $\mathbf{E}$ is determined by equation:
\begin{equation}
4\pi \rho_{bond}=div~\mathbf{E}.\label{divgg}
\end{equation}
It was calculated before \cite{V-01} that the equilibrium equation
Eq.({\ref{gen31}) can be realized at condition of very high
temperature and density of plasma only, i.e. in a central region of
a star. As the gradient of pressure is absent at the equilibrium
condition (Eq.({\ref{gen31})), it is not difficult to conclude that
a core with a constant particle (electron) density in plasma
\cite{V-01}
\begin{equation}
{\eta_\star}=\frac{16}{9\pi}\frac{(Z+1)^3}{r_B^3},\label{eta}
\end{equation}
 at constant temperature
\begin{equation}
\mathbb{T}=\biggl(\frac{10}{\pi^4}\biggr)^{1/3}(Z+1)\frac{\hbar
c}{kr_B}\approx 2  \cdot 10^7(Z+1)~K.\label{Ts}
\end{equation}
must be placed in centrum of a star.

Its mass must be equal to \cite{V-01}
\begin{equation}
{\mathbb{M}}=1.5^6\biggl(\frac{10}{\pi^{3}}\biggr)^{1/2}
\biggl(\frac{\hbar c} {Gm_p^2}\biggr)^{3/2}
\biggl(\frac{Z}{A}\biggr)^2 m_p \approx 6.47~ M_{Ch}~
\biggl(\frac{Z}{A}\biggr)^2\label{Ms},
\end{equation}
where $M_{Ch}=\biggl(\frac{\hbar
c}{Gm_p^2}\biggr)^{3/2}m_p=3.71\cdot 10^{33} g$ is the Chandrasekhar
mass, $A$ and $Z$ are the averaged mass and charge numbers of atomic
nuclei from which the plasma is composed, $r_B$ is the Bohr radius,
$m_p$ is the proton mass. The radius of a star core
\begin{eqnarray}
\mathbb{R}= \frac{(3/2)^3}{2}
\biggl(\frac{10}{\pi}\biggr)^{1/6}\biggl(\frac{\hbar c}{G
m_p^2}\biggr)^{1/2} \frac{r_B}{(Z+1){A/Z}}.\label{Rs}
\end{eqnarray}
must be equal approximately to 1/10 of an external radius $R_0$ of a
star. It can be seen, that all main parameters of a star core are
depending on a chemical composition ($A/Z$ and  $Z+1$) of its plasma
only.

A substance in region above a star core must be three order of
magnitude more rarefied. This region can be named as an atmosphere
of a star.

A testing of the validity of a fundamental postulate of a theory is
a standard procedure which was developed by the scientific community
since G.Galileo: the laws must by formulated by means of standard
mathematical methods on the base of the tested postulate. Then one
must check empirically that the Nature "obeys" these laws and as
result to confirm the validity of the fundamental postulate. In our
case on a base of the postulate (Eq.({\ref{eP}})) (or in a more
convenient form Eq.({\ref{gen31}})), the theoretical conclusions
about main characteristics of a star core have been deduced
(Eq.({\ref{eta}}),Eq.({\ref{Ts}}),Eq.({\ref{Ms}}), Eq.({\ref{Rs}})).
But the astronomers cannot measure the internal characteristic of
stars directly. They can measure under some condition the integral
and surface characteristics of stars - masses, radii, surface
temperatures, luminosities. To have the possibility to compare
theoretical predictions with measured characteristics of stars, we
must calculate corresponding parameters of a star atmosphere. The
solution is possible if one considers a stationary stable state of a
star as its equilibrium one. A hot star generates an energy
continuously and radiates it from its surface. This radiation is
non-equilibrium relatively to a star, but it can be stationary
stable. The substance of a star can be considered to have an
equilibrium state in adiabatic conditions because the existing
exchange of energy between the radiation subsystem and the subsystem
of substance is permanent and doesn't induce an alteration of state
of the substance. Hence the description of a state of a star
substance can be based on a consideration of an equilibrium
condition of  a hot plasma or, in the first approximation, on a
consideration of an equilibrium of  an ideal gas in adiabatic
conditions.

\section{The equilibrium state of substance of a star interior and stellar masses}

An ideal gas  in a volume  without gravitation exists in equilibrium
state at a pressure equalization, i.e. at an equalization of
temperature $T$ and particle density $n$ over volume. The constancy
of chemical potential $\mu$ is mandatory requirement of this
equilibrium state.

If a different parts of a system have different temperatures in an
equilibrium state, the mandatory requirement of equilibrium adds  up
to \cite{LL}:
\begin{equation}
\frac{\mu}{kT}=const
\end{equation}
As thermodynamical (statistical) part of chemical potential of an
ideal gas \cite{LL}:
\begin{equation}
\mu_T= kT~ln \biggl[\frac{n}{2}\biggl(\frac{2\pi \hbar^2}{m
kT}\biggr)^{3/2}\biggr],\label{muB}
\end{equation}
the density of ideal gas in equilibrium state must depend on
temperature
\begin{equation}
n\sim T^{3/2}.\label{32}
\end{equation}
The chemical potential of a system at gravity action \cite{LL}
\begin{equation}
\mu=\mu_T + E_{G}
\end{equation}
where $E_{G}$ is the gravitational energy of particle. Hence the
equilibrium state needs a fulfilment of additional condition:
\begin{equation}
\frac{GM_r m}{r kT_r}=const
\end{equation}
(where $m$ is the mass of a particle, $M_r$ is the mass of star
substance inside spherical volume with radius $r$, $T_r$ is the
temperature on this surface) or
\begin{equation}
M_r \sim r kT_r.\label{mtr}
\end{equation}
If we suppose that temperature inside star behaves as a power with
exponent $x$, its value on the radius $r$
\begin{equation}
T_r=T_\star\biggl(\frac{\mathbb{R}}{r}\biggr)^x\label{Tr}
\end{equation}
and  accordingly the particle density
\begin{equation}
n_r=\eta_\star\biggl(\frac{\mathbb{R}}{r}\biggr)^{3x/2}.\label{nr}
\end{equation}
From Eq.({\ref{mtr}), at equalizing of the exponent at $r$ in the
left and right parts, one obtains $x=4$, i.e.
\begin{equation}
n_r=\eta_\star\biggl(\frac{\mathbb{R}}{r}\biggr)^6\label{nr6}
\end{equation}
and
\begin{equation}
T_r=\mathbb{T}\biggl(\frac{\mathbb{R}}{r}\biggr)^4\label{tr}
\end{equation}
Hence the mass of star atmosphere
\begin{equation}
M_{A} = 4\pi \int_{\mathbb{R}}^{R_0} m' \eta_\star
\biggl(\frac{\mathbb{R}}{r}\biggr)^6 r^2 dr\approx \frac{4\pi}{3}m'
\eta_\star \mathbb{R}^3=\mathbb{M}\label{ma}
\end{equation}
is equal to mass of its core (accurate to
$\frac{\mathbb{R}^3}{R_0^3}\approx 10^{-3}$), where $m'$ is the mass
of star plasma related to one electron.

This claim is in a good agreement with measurement data of Sun
properties. The core mass of the Sun was calculated before
\cite{VAP-Sun} on a base of data of measurement of natural
frequencies of seismical sunny oscillations \cite{GOLF}. According
to this data the core mass of the Sun equals to $9.6~10^{32}~g$ (it
is $0.48$ mass of the Sun), i.e. for the Sun $M_A\approx \mathbb{M}$
really.

As a result the full mass of a star
\begin{equation}
M = 2 \mathbb{M}\approx 12.9~
M_{Ch}~\biggl(\frac{Z}{A}\biggr)^2\label{2mstar}
\end{equation}
depends of the ratio $A/Z$ only, and for comparison this estimation
with measurement data we need the chemical composition of distant
stars, which is unknown. But some predictions in this direction are
possible. At first, there must be no stars with masses exceeding the
Sun mass more than one and a half orders, because it corresponds to
the limiting mass of stars consisting from hydrogen with $A/Z = 1$.
Secondly, though the neutronization process makes neutron-excess
nuclei stable, there is no reason to suppose that stars with $A/Z$
which are essentially larger than few units (and with mass almost in
hundred times less than hydrogen stars) can exist. Thus, the theory
predicts that the whole mass spectrum of stellar masses must be
placed in the interval from few tenth up to approximately 25 solar
masses. These predictions are verified by measurements quite
exactly. The mass distribution of binary stars is shown on
Fig.{\ref{stars}} \cite{Heintz}. (Using of this data is caused by
the fact that only the measurement of parameters of the binary star
rotation gives the possibility to determine their masses with
satisfactory accuracy). Besides, one can see the presence of
separate pikes for stars with $A/Z = 3; 4; 5...$ and with $A/Z =
3/2$ on Fig.{\ref{stars}}. The stars with $A/Z = 2$  are observed
too, but they don't form a separate peak on this histogram.
\begin{figure}
\begin{center}
\centering
\includegraphics[scale=0.7]{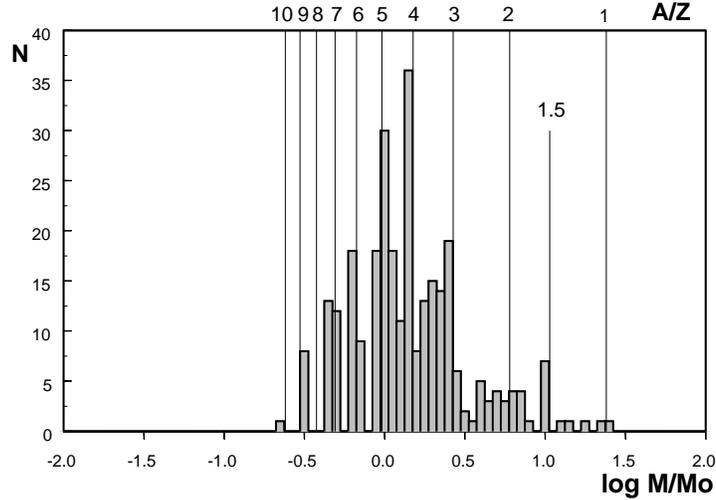}
%
\caption{The mass distribution of binary stars \cite{Heintz}. On
abscissa, the logarithm of the star mass over the Sun mass is shown.
Solid lines mark masses agree with selected values of $A/Z$ from
Eq.({\ref{2mstar}})} {\label{stars}}
\end{center}
\end{figure}
It was mentioned above, that the chemical composition of the sunny
core can be calculated  \cite{VAP-Sun} on  base of the measurement
data of natural seismic oscillations of the sunny surface
\cite{GOLF}. This calculation shows that the core of the Sun is
basically consisting from hellium-10 (A/Z=5). This is in full
agreement with its position on histogram {Fig.(\ref{stars})}.

\section{The mass-radius ratio}
In according with the virial theorem \cite{LL,VL}, the full energy
of a particle inside a star must be equal to its kinetic energy with
sign minus:
\begin{equation}
E=-\frac{3}{2}kT
\end{equation}

In this case the heat capacity at a constant volume (per particle
over Boltzman constant $k$) by definition is
\begin{equation}
c_v=-\frac{3}{2}
\end{equation}
The negative heat capacity of stellar substance is not surprising.
It is a known fact which was discussed in Landau-Lifshitz course
\cite{LL}. The own heat capacity of each particle of the star
substance is positive. One obtains the negative heat capacity taking
into account the gravitational interaction between particles. The
definition of the heat capacity of an ideal gas particle at
permanent pressure \cite{LL}
\begin{equation}
c_p=c_v+1
\end{equation}
gives
\begin{equation}
c_p=-\frac{1}{2}
\end{equation}
and the adiabatic exponent
\begin{equation}
\widetilde{\gamma}=\frac{c_p}{c_v}=\frac{1}{3}.
\end{equation}
Supposing that a stellar atmosphere is adiabatic, we can use the
Poisson adiabat  definition  \cite{LL}
\begin{equation}
\widehat{P}V_0^{\widetilde{\gamma}}=const
\end{equation}
and obtain
\begin{equation}
\widehat{P}R_0=const.
\end{equation}
From pressure averaged over star volume
\begin{equation}
\widehat{P}\approx\frac{GM^2}{R_0^4}
\end{equation}
we obtain desired relation between mass and radius of a star:
\begin{equation}
\frac{M^2}{R_0^3}=const\label{rm23}
\end{equation}
Simultaneously the observational data of  masses, of radii and their
temperatures was obtained by astronomers for close binary stars. For
convenience of readers, the data of these measurements for 100 stars
from 50 close pairs are gathered in Table 1 in Appendix of this
article. The dependence of radii of these stars over these masses is
shown on Fig.{\ref{RM}} in twice logarithmic scale. The solid line
shows the result of fitting of measurement data $R\sim M^{0.68}$. It
is close to theoretical dependence $R\sim M^{2/3}$~(Eq.{\ref{rm23}})
which is shown by the dotted line.
\begin{figure}
\begin{center}
\includegraphics[scale=0.7]{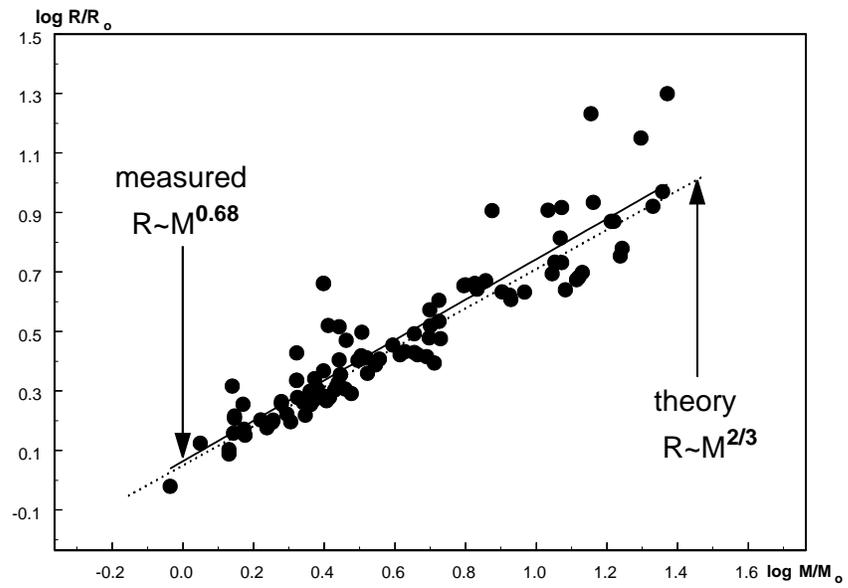}
\caption{The dependence of radii of stars over these masses. The
radii  of stars are normalized to the sunny radius, the stars masses
are normalized to the mass of the Sum. The data are shown in twice
logarithmic scale. The solid line shows the result of fitting of
measurement data ($R\sim M^{0.68}$). The theoretical dependence
($R\sim M^{2/3}$~(Eq.{\ref{rm23}})) is shown by the dotted line.}
{\label{r-m}}
\end{center}
\end{figure}

\section{The mass-temperature and mass-luminosity relations}

According to Eq.{\ref{tr}} the temperature on the star surface
\begin{equation}
T_0=\mathbb{T}\biggl(\frac{\mathbb{R}}{R_0}\biggr)^4.\label{tr1}
\end{equation}
Taking into account the characteristics of the star core
(Eq.{\ref{Ts}})-(Eq.{\ref{Rs}}) and (Eq.{\ref{rm23}}), we obtain a
relation for main parameters of star atmosphere
\begin{equation} R_0 T_0\sim
\frac{1}{(1+Z)^3}\label{TR2}
\end{equation}
If we suppose as above that a star atmosphere is in adiabatic
conditions, its averaged parameters can be related by the Poisson
adiabat equation \cite{LL}:
\begin{equation}
\widehat{T}^{\widetilde{\gamma}}
\widehat{P}^{1-\widetilde{\gamma}}=const
\end{equation}
The averaged temperature of a star atmosphere can be obtained by
integrating over its volume:
\begin{equation}
\widehat{T}\sim T_0 \biggl(\frac{R_0}{\mathbb{R}}\biggr).\label{TR3}
\end{equation}
Averaged parameters of substance inside a star atmosphere can be
described by the ideal gas law
\begin{equation}
\widehat{P}=k\widehat{T}~\widehat{n},
\end{equation}
where averaged particle density into atmosphere
\begin{equation}
\widehat{n}\approx \frac{N_A}{R_0^3}.
\end{equation}
Taking into account (Eq.{\ref{ma}}) and (Eq.{\ref{Ms}}), full number
particles into atmosphere
\begin{equation}
N_A\sim (A/Z)^{-3/2}
\end{equation}
we obtain
\begin{equation}
T_0\sim R_0^{7/8},
\end{equation}
or considering (Eq.{\ref{rm23}})
\begin{equation}
T_0\sim \mathbb{M}^{7/12}.\label{tm}
\end{equation}
The dependence of the temperature on the star surface over the star
mass for the same set of data as before on Fig.(\ref{RM}). Here the
temperatures of stars are normalized to the sunny surface
temperature (5875~C), the stars masses are normalized to the mass of
the Sum. The data are shown in twice logarithmic scale. The solid
line shows the result of fitting of measurement data ($T_0\sim
M^{0.59}$). The theoretical dependence $T_0\sim
M^{7/12}$~(Eq.{\ref{tm}}) is shown by the dotted line.} {\label{RM}}
\begin{figure}
\begin{center}
\includegraphics[scale=0.7]{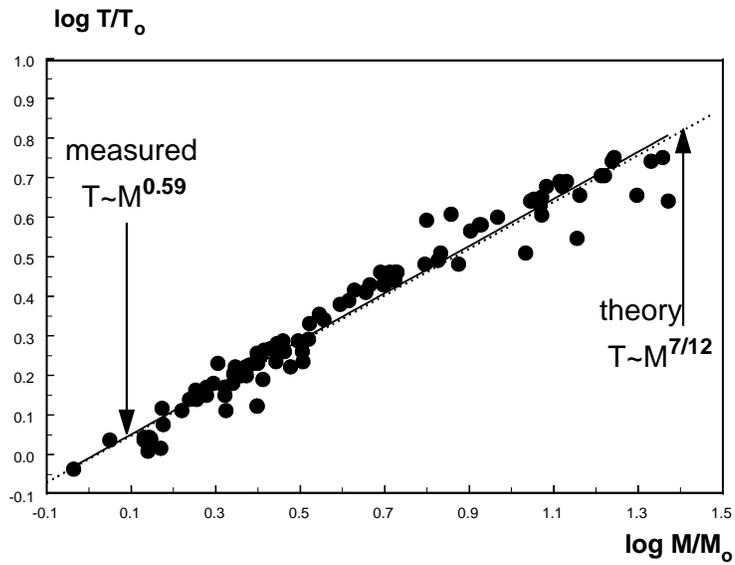}
\caption{The dependence of the temperature on the star surface over
the star mass for the same set of data as before on Fig.(\ref{RM}).
The temperatures of stars are normalized to the sunny surface
temperature (5875~C), the stars masses are normalized to the mass of
the Sum. The data are shown in twice logarithmic scale. The solid
line shows the result of fitting of measurement data ($T_0\sim
M^{0.59}$). The theoretical dependence $T_0\sim
M^{7/12}$~(Eq.{\ref{tm}}) is shown by the dotted line.}
{\label{t-m}}
\end{center}
\end{figure}

The luminosity of a star
\begin{equation}
L_0\sim R_0^2 T_0^4.
\end{equation}
Taking into account (Eq.{\ref{rm23}}) and (Eq.{\ref{tm}}) it can be
expressed as
\begin{equation}
L_0\sim \mathbb{M}^{11/3}\sim \mathbb{M}^{3.67}\label{lm}
\end{equation}
This dependence is shown in Fig.(\ref{l-m})
\begin{figure}
\begin{center}
\includegraphics[scale=0.7]{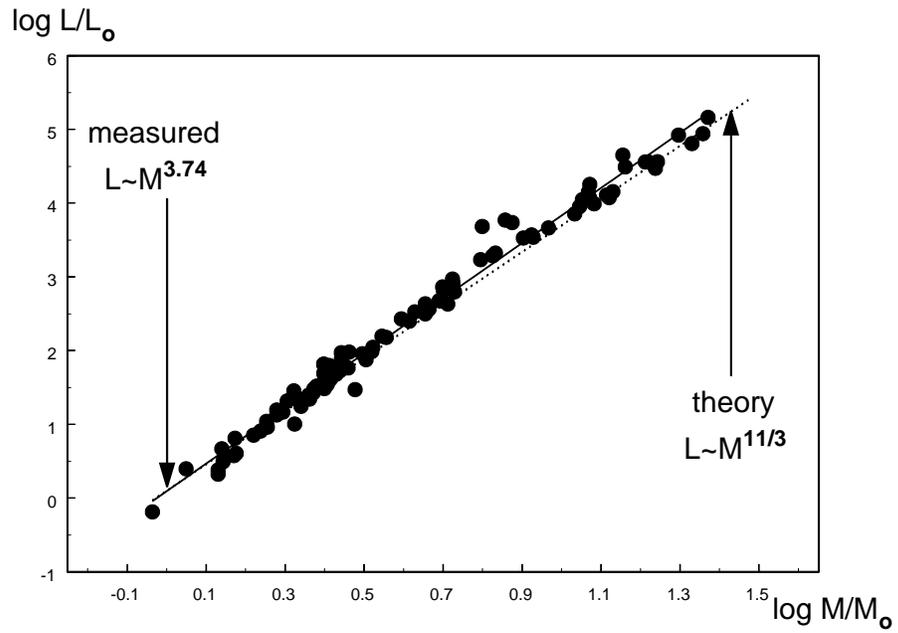}
\caption{The dependence of star luminosity over the star mass for
the same set of data as before on Fig.(\ref{r-m}) and
Fig.(\ref{t-m}). The luminosities are normalized to the luminosity
of the Sun, the stars masses are normalized to mass of the Sum. The
data are shown in twice logarithmic scale. The solid line shows the
result of fitting of measurement data $L\sim M^{3.74} $. The
theoretical dependence $L\sim M^{11/3}$~(Eq.{\ref{lm}}) is shown by
the dotted line.} {\label{l-m}}
\end{center}
\end{figure}
\section{Conclusion}
It can be seen that the theoretically obtained dependencies
describing the star mass distribution and mass-radius,
mass-temperature and mass-luminosity relations are in a good
agreement with measurement data. These theoretical results  together
with dependencies explaining the velocity of the periastron rotation
of close binary stars \cite{VAP-Per} and explaining the spectrum of
seismical oscillations of the Sun \cite{VAP-Sun} are composing a
practically full set of theoretical predictions that can by checked
by a comparison with the measurement data. Their accordance
unambiguously and confidently shows that there are cores in central
regions of stars. Here the gravity force is balanced by a force with
electric nature. This accordance confirms the validity of basic
postulate Eq.({\ref{eP}}).

The comparison of theoretical dependencies mass-radius,
mass-temperature, mass-luminosity and measurement data became
technical possible when information about main parameters of close
binary stars was obtained, because it keeps the set of data about
masses, radii and temperatures of stars simultaneously. This
information is the result of measurements of  a lot of astronomers
from a lot of countries. At last time the summary table with these
data was gathered by Khaliulilin Kh.F. in his dissertation (in
Russian) which has unfortunately a restricted access. With his
agreement and for readers convenience, we place this table in
Appendix.


\section{Appendix}
\hspace{-5.5cm}
{\scriptsize
\begin{tabular}
{||r|l|r|r|r|r|r|r|r|r|c||}\hline \hline
   & & & & & & & & & & \\
  N & Name of star & U & P & $M_1/M_{\odot}$ & $M_2/M_{\odot}$ &  $R_1/R_{\odot}$ & $R_2/R_{\odot}$& $T_1$ & $T_2$ &References\\
& &period of&period of &mass of&mass of & radius of  &radius of  &temperature &temperature & \\
&
&apsidal&elipsodal &component 1,&component 2,&1 component& 2 component&of&of& \\
& &rotation,& rotation,&in&in&in &in &1 component,&2 component, & \\
& &years & days&the Sun mass&the Sun mass &the Sun radius &the Sun
radius &K &K&
\\\hline
  1 & BW Aqr & 5140 & 6.720 & 1.48 & 1.38 &  1.803 & 2.075 & 6100 & 6000 & 1,2\\
  2 & V 889 Aql & 23200 & 11.121 & 2.40 & 2.20& 2.028 & 1.826 & 9900 & 9400 & 3,4 \\
  3 & V 539 Ara & 150 & 3.169 & 6.24 & 5.31 &  4.512 & 3.425 & 17800 & 17000  &5,12,24,67 \\
  4 & AS Cam & 2250 & 3.431 & 3.31 & 2.51 &   2.580 & 1.912 & 11500 & 10000 & 7,13\\
  5 & EM Car & 42 & 3.414 & 22.80 & 21.40 &   9.350 & 8.348 & 33100 & 32400 & 8\\
  6 & GL Car & 25 & 2.422 & 13.50 & 13.00 &   4.998 & 4.726 & 28800 & 28800 & 9\\
  7 & QX Car & 361 & 4.478 & 9.27 & 8.48 &   4.292 & 4.054 & 23400 & 22400  & 10,11,12\\
  8 & AR Cas & 922 & 6.066 & 6.70 & 1.90 &   4.591 & 1.808 & 18200 & 8700 & 14,15\\
  9 & IT Cas & 404 & 3.897 & 1.40 & 1.40 &   1.616 & 1.644 & 6450 & 6400 & 84,85\\
  10 & OX Cas & 40 & 2.489 & 7.20 & 6.30 &   4.690 & 4.543 & 23800 & 23000 & 16,17\\
  11 & PV Cas & 91 & 1.750 & 2.79 & 2.79 &   2.264 & 2.264 & 11200 & 11200  & 18,19\\
  12 & KT Cen & 260 & 4.130 & 5.30 & 5.00 &  4.028 & 3.745 & 16200 & 15800 & 20,21\\
  13 & V 346 Cen & 321 & 6.322 & 11.80 & 8.40  & 8.263 & 4.190 & 23700 & 22400 & 20,22\\
  14 & CW Cep & 45 & 2.729  & 11.60 & 11.10 &  5.392 & 4.954 & 26300 & 25700  & 23,24\\
  15 & EK Cep & 4300 & 4.428 & 2.02 & 1.12 &  1.574 & 1.332 & 10000  & 6400  &25,26,27,6\\
  16 & $\alpha$ Cr B & 46000 & 17.360 & 2.58 & 0.92  & 3.314 & 0.955 & 9100 & 5400 & 28,29\\
  17 & Y Cyg & 48 & 2.997 & 17.50 & 17.30 &   6.022 & 5.680 & 33100 & 32400 & 23,30\\
  18 & Y 380 Cyg & 1550 & 12.426 & 14.30 & 8.00   & 17.080 & 4.300 & 20700 & 21600 &31 \\
  19 & V 453 Cyg & 71 & 3.890 & 14.50 & 11.30 &  8.607 & 5.410 & 26600 & 26000 &17,32,33\\
  20 & V 477 Cyg & 351 & 2.347 & 1.79 & 1.35 &   1.567 & 1.269 & 8550 & 6500 & 34,35\\
  21 & V 478 Cyg & 26 & 2.881 & 16.30 & 16.60 &  7.422 & 7.422 & 29800 & 29800 & 36,37\\
  22 & V 541 Cyg & 40000 & 15.338 & 2.69 & 2.60  & 2.013 & 1.900 & 10900 & 10800 & 38,39\\
  23 & V 1143 Cyg & 10300 & 7.641 & 1.39 & 1.35  & 1.440 & 1.226 & 6500 & 6400 & 40,41,42\\
  24 & V 1765 Cyg & 1932 & 13.374 & 23.50 & 11.70   & 19.960 & 6.522 & 25700 & 25100 & 28\\
  25 & DI Her & 29000 & 10.550  & 5.15 & 4.52 &   2.478 & 2.689 & 17000 & 15100 & 44,45,46,47\\
  26 & HS Her & 92 & 1.637  & 4.25 & 1.49 &   2.709 & 1.485 & 15300 & 7700 & 48,49\\
  27 & CO Lac & 44 & 1.542 & 3.13 & 2.75 &   2.533 & 2.128 & 11400 & 10900 & 50,51,52\\
  28 & GG Lup & 101 & 1.850 & 4.12 & 2.51 &  2.644 & 1.917 & 14400 & 10500 & 17\\
  29 & RU Mon & 348 & 3.585 & 3.60 & 3.33 &  2.554 & 2.291 & 12900 & 12600  & 54,55\\
  30 & GN Nor & 500 & 5.703 & 2.50 & 2.50 &  4.591 & 4.591 & 7800 & 7800  & 56,57\\
  31 & U Oph & 21 & 1.677 & 5.02 & 4.52 &   3.311 & 3.110 & 16400 & 15200 & 53,58,37\\
  32 & V 451 Oph & 170 & 2.197 & 2.77 & 2.35  & 2.538 & 1.862 & 10900 & 9800 & 59,60\\
  33 & $\beta$ Ori & 228 & 5.732 & 19.80 & 7.50   & 14.160 & 8.072 & 26600 & 17800 & 61,62,63\\
  34 & FT Ori & 481 & 3.150 & 2.50 & 2.30 &   1.890 & 1.799 & 10600 & 9500 & 64\\
  35 & AG Per & 76 & 2.029 & 5.36 & 4.90 &  2.995 & 2.606 & 17000 & 17000 &23,24\\
  36 & IQ Per & 119 & 1.744 & 3.51 & 1.73 &  2.445 & 1.503 & 13300 & 8100 &65,66\\
  37 & $\zeta$ Phe & 44 & 1.670 & 3.93 & 2.55  & 2.851 & 1.852 & 14100 & 10500 &11,67 \\
  38 & KX Pup & 170 & 2.147 & 2.50 & 1.80 &   2.333 & 1.593 & 10200 & 8100 & 21\\
  39 & NO Pup & 37& 1.257 & 2.88 & 1.50 &   2.028 & 1.419 & 11400 & 7000 &11,69\\
  40 & VV Pyx & 3200 & 4.596 & 2.10 & 2.10  & 2.167 & 2.167 & 8700 & 8700 &70,71\\
  41 & YY Sgr & 297 & 2.628 & 2.36 & 2.29 & 2.196 & 1.992 & 9300 & 9300 & 72\\
  42 & V 523 Sgr & 203 & 2.324 & 2.10 & 1.90   & 2.682 & 1.839 & 8300 & 8300 &73 \\
  43 & V 526 Sgr & 156 & 1.919 & 2.11 & 1.66 &   1.900 & 1.597 & 7600 & 7600 &74\\
  44 & V 1647 Sgr & 592 & 3.283 & 2.19 & 1.97 &  1.832 & 1.669 & 8900 & 8900 & 75\\
  45 & V 2283 Sgr & 570 & 3.471 & 3.00 & 2.22 &  1.957 & 1.656 & 9800 & 9800 & 76,77\\
  46 & V 760 Sco & 40 & 1.731 & 4.98 & 4.62 &   3.015 & 2.642 & 15800 & 15800 & 78\\
  47 & AO Vel & 50 & 1.585 & 3.20 & 2.90 &   2.623 & 2.954 & 10700 & 10700 & 79\\
  48 & EO Vel & 1600 & 5.330 & 3.21 & 2.77 &  3.145 & 3.284 & 10100 & 10100 & 21,63\\
  49 & $\alpha$ Vir & 140 & 4.015 & 10.80 & 6.80  & 8.097 & 4.394 & 19000 & 19000 &80,81,68 \\
  50 & DR Vul & 36 & 2.251 & 13.20 & 12.10 &   4.814 & 4.369 & 28000 & 28000 & 82,83\\ \hline\hline
\end{tabular}}


\clearpage
\begin{thebibliography}{7}
\bibitem {V-01}    Vasiliev B.V. - Nuovo Cimento B, 2001, v.116, pp.617-634.
\bibitem {LL}    Landau L.D. and Lifshits E.M. - Statistical Physics,1980, vol.1,3rd edition,Oxford:Pergamon.
\bibitem {VL}    Vasiliev B.V. and Luboshits V.L. $\textit{Physics-Uspekhi}, \bf{ 37}$ (1994) 345.
\bibitem {VAP-Per}    Vasiliev B.V. - Astro-ph/0405297
\bibitem {GOLF}    Solar Physics, vol.175/2, (http://sohowww.estec.esa.nl/gallery/GOLF)
\bibitem {VAP-Sun}    Vasiliev B.V. - Astro-ph/0409491
\bibitem {Heintz}    Heintz W.D. $\textit{Double stars}$ (Geoph. and Astroph. monographs. vol.15, D.Reidel Publ. Corp.)
1978.
\end{thebibliography}

\clearpage

\begin{thebibliography}{85}


\bibitem {KhKo}   Khaliulilin Kh.F. and Kozyreva V.S.

\textit{Apsidal motion in the eclipsing binary system of BW Aqr}

Astrophys. and Space Sci., {\bf{120}}  (1986) 9-16.



\bibitem {Imbert}   Imbert M.

\textit{Photoelectric radial velosities of eclipsing binaries. IV.
Orbital elements of BW Aqr},

Astron.Astrophys.Suppl., {\bf {69}} (1987) 397-401.





\bibitem {KhKh}   Khaliulilin Kh.F. and Khaliulilina A.I.

\textit{Fotometricheskoe issledovanie zatmenno-dvoinoi sistemy s
relativistskim vrasheniem orbity V889 Aql,}

Astronom.zh.,{\bf{66}}(1989)76-83 (in Russian).




\bibitem {KhKh87}
Khaliulilin Kh.F. and Khaliulilina A.I.

\textit{K probleme vrashenia linii apsid v zatmennoi sisteme V889
Aql},

Astron.cirk., {\bf N{1486}}  (1987) 5-7 (in Russian).




\bibitem {Clausen}
Clausen J.V.

\textit{V 539 Arae: first accurate dimensions of eclipsing
binaries},

Astron.Astrophys., {\bf {308}} (1996) 151-169.




\bibitem {LavLav}
Lavrov M.I. and Lavrova N.V.

\textit{Revisia elementov fotometricheskoi orbity EK Cep},

Astron.cirk. {\bf {971}} (1977) 3-4 (in Russian).




\bibitem {KhKo94}   Khaliulilin Kh.F. and Kozyreva V.S.
\textit{Apsidal motion in the eclipsing binary AS Cam},

Astrophys. and Space Sci., {\bf{120}}  (1994) 115-122.




\bibitem {Anders89}   Andersen J. and Clausen J.V.,
\textit{Absolute dimensions of eclipsing binaries.XV. EM Cainae},

Astron.Astrophys. {\bf{213}} (1989) 183-194.




\bibitem {Gemenez}   Gemenez A. and Clausen J.V.,

\textit{Four-color photometry of eclipsing binaries. XXIIA.
Photometric elements and apsidal motion of GL Cainae},

Astron.Astrophys. {\bf{161}} (1986) 275-286.




\bibitem {Anders83}   Andersen J., Clausen J.V., Nordstrom B. and
Reipurth B.,

\textit{Absolute dimensions of eclipsing binaries.I. The early-type
detached system QX Cainae},

Astron.Astrophys. {\bf{121}} (1983) 271-280.




\bibitem {Gemenez86b}   Gemenez A., Clausen J.V. and Jensen K.S.

\textit{Four-color photometry of eclipsing binaries. XXIV. Aspidal
motion of QX Cainae, $\xi$ Phoenicis and NO Puppis},

Astron.Astrophys. {\bf{159}} (1986) 157-165.




\bibitem {DeGreve}   De Greve J.P.

\textit{Evolutionary models for detached close binaries: the system
Arae and QX Cainae},

Astron.Astrophys. {\bf{213}} (1989) 195-203.




\bibitem {Malony}   Malony F.P., Guinan E.F. and Mukherjec J.

\textit{Eclipsing binary star as test of gravity theories}

Astron.J. {\bf{102}} (1991) 256-261.



\bibitem {Mossak92}   Mossakovskaya L.V.

\textit{New photometric elements of AR Cas, an eclipsing binary
system with apsidal motion}

Astron. and Astroph. Trans. {\bf{2}} (1992) 163-167.



\bibitem {Haffer62}   Haffer C.M. and Collins G.M.

\textit{Computation of elements of eclipsing binary stars by
high-speed computing machines}

Astroph.J.Suppl., {\bf{7}} (1962) 351-410.





\bibitem {Crink89}   Crinklaw G. and Etzel P.

\textit{A photometric analisis of the eclipsing binary OX
Cassiopeiae}

Astron.J. {\bf{98}} (1989) 1418-1426.




\bibitem {Claret93}   Claret A. and Gimenez A.

\textit{The aspidal motion test of the internal stellar structure:
comparision between theory and observations}

Astron.Astroph. {\bf{277}} (1993) 487-502.




\bibitem {Wolf95a}   Wolf M.

\textit{Aspidal motion in the eclipsing binary PV Cassiopeiae}

Monthly Not.Roy.Soc. {\bf{286}} (1995) 875-878.




\bibitem {Popper87}   Popper D.M.

\textit{Rediscussion of eclipsing binaries.XVII.The detached early A
type binaries PV Cassiopeae and WX Cephei}

Astron.J. {\bf{93}} (1987) 672-677.



\bibitem {LavLav85}
Lavrov M.I. and Lavrova N.V.

\textit{Revisia fotometrichestih elementov u zatmennyh dvoinyh
sistem s ekscentricheskimi orbitami.2.KT Cen}

Trudy Kaz.Gor.AO {\bf {49}} (1985) 18-24 (in Russian).



\bibitem {Soder75}   Soderhjelm S.

\textit{Observations of six southern eclipsing  binaries for apsidal
motion}

Astron.Astroph.Suppl.Ser {\bf{22}} (1975) 263-283.



\bibitem {Gemenez86a}   Gemenez A., Clausen J.V. and Anderson J.

\textit{Four-color photometry of eclipsing binaries. XXIA.
Photometric  analysis and aspidal motion study of V346 Centauri},

Astron.Astrophys. {\bf{160}} (1986) 310-320.



\bibitem {Gemenez87}   Gemenez A., Chun-Hwey Kim and Il-Seong Nha

\textit{Aspidal motion in the early-type eclipsing binaries CW
Cephei, Y Cyg and AG Per}

Montly .Not.Roy.Astron.Soc. {\bf{224}} (1987) 543-555.



\bibitem {Boz}   Bocula R.A.

\textit{Peresmotr elementov fotometricheskoi orbity zatmennyh sistem
CW Cep, V 539 Ara, AG Per, AR Aur, RS Cha, ZZ Boo.}

Peremennye zvezdy{\bf{21}} (1983) 851-859 (in Russian).



\bibitem {Kh83}   Khaliulilin Kh.F.

\textit{Relativistskoe vrashenie orbity zatmennoi dvoinoi sistemy EK
Cep}

Astron.zh.{\bf{60}} (1983) 72-82 (in Russian).



\bibitem {Tomkin}   Tomkin J.

\textit{Secondaries of eclipsing binary. V. EK Cephei}

Astroph.J. {\bf{271}} (1983) 717-724.



\bibitem {Claret}   Claret A., Gemenez A. and Martin E.L.

\textit{A test case of stellar evolution the  eclipsing binary EK
Cephei}

Astron.Astroph. {\bf{302}} (1995) 741-744.



\bibitem {Volk}   Volkov I.M.

\textit{The discovery of apsidal motion in the  eclipsing binary
system $\alpha$ Cr B}

Inf.Bull.Var.Stars N3876,(1993) 1-2.



\bibitem {Quiroga}   Quiroga R.J., van L.P.R.

\textit{Angular momenta in binary systems}

Astroph.Space Sci. {\bf{146}} (1988) 99-137.



\bibitem {Hill95}   Hill G. and Holmgern D.E.

\textit{Studies of early-type varieble stars}

Asrton.Astroph.{\bf{297}} (1995) 127-134.



\bibitem {Hill84}   Hill G. and Batten A.H.

\textit{Studies of early-type varieble stars.III. The orbit and
physical dimensions for V 380 Cygni}

Asrton.Astroph.{\bf{141}} (1984) 39-48.



\bibitem {Zak}   Zakirov M.M.

\textit{Ob apsidalnom dvizhenii v dvoinoi sisteme  V 453 Cyg}

Astron.cirk.N1537,21 (in Russian).



\bibitem {Kar}   Karetnikov V.G.

\textit{Spectral investigation of eclipsing binary stars at the
stage of mass exchange}

Publ.Astron.Inst.Czech.{\bf{70}} (1987) 273-276.



\bibitem {Mossak87}   Mossakovskaya L.V. and Khaliulilin Kh.F.

\textit{Prichina anomalnogo apsidalnogo dvizhenia v sisteme V 477
Cyg}

Astron.cirk.N1536, 23-24 (in Russian).



\bibitem {Gemenez92}   Gemenez A. and Quintana J.M.

\textit{Apsidal motion and revised photometry elements of the
eccentric eclipsing binary V 477 Cyg}

Astron.Astrophys. {\bf{260}} (1992) 227-236.



\bibitem {Mossak96}   Mossakovskaya L.V. and Khaliulilin Kh.F.

\textit{Vrashenie linii apsid v sisteme  V 478 Cyg}

Pisma v Astron.zh.{\bf{22}} (1996) 149-152.



\bibitem {Popper91}   Popper D.M. and Hill G.

\textit{Rediscussion of eclipsing binaries.XVII.Spectroscopic orbit
of OB system with a cross-correlation procedure}

Astron.J. {\bf{101}} (1991) 600-615.



\bibitem {Kh85}   Khaliulilin Kh.F.

\textit{The unique eclipsing binary system V 541 Cygni with
relativistic apsidal motion}

Astrophys.J. {\bf{229}} (1985) 668-673.



\bibitem {Lin}   Lines R.D.,Lines H., Guinan E.F. and Carroll

\textit{Time  of minimum determination  of eclipsing binary V 541
Cygni}

Inf.Bull.Var.Stars N3286,1-3.



\bibitem {Kh83b}   Khaliulilin Kh.F.

\textit{Vrashenie linii apsid v zatmennoi sisteme V 1143 Cyg}

Asrton. cirk.N1262,1-3 (in Russian).



\bibitem {Anders87}   Andersen J., Garcia J.M., Gimenes A. and
Nordstom B.

\textit{Absolute dimension of eclipsing binaries.X. V1143 Cyg}

Astron.Astrophys. {\bf{174}} (1987) 107-115.



\bibitem {Burns}   Burns J.F., Guinan E.F. and Marshall J.J.

\textit{New apsidal motion determination of eccentric  eclipsing
binary V 1143 Cyg}

Inf.Bull.Var.Stars N4363,1-4.



\bibitem {Hill85}   Hill G. and Fisher W.A.

\textit{Studies of early-type varieble stars.II. The orbit and
masses ofHR 7551}

Astron.Astrophys. {\bf{139}} (1985) 123-131.



\bibitem {Mart80}   Martynov D.Ya. and  Khaliulilin Kh.F.

\textit{On the relativistic motion of periastron in the eclipsing
binary system DI Her}

Astrophys.and Space Sci. {\bf{71}} (1980) 147-170.



\bibitem {Popper82}   Popper D.M.

\textit{Rediscussion of eclipsing binaries.XVII. DI Herculis, a
B-tipe system with an accentric orbit}

Astron.J. {\bf{254}} (1982) 203-213.



\bibitem {Mart97}   Martynov D.Ia. è Lavrov M.I.

\textit{Revizia elementov fotometricheskoi orbity i skorosti
vrashenia linii apsid u zatmennoi dvoinoi sistemy DI Her}

Pisma v Astron.Zh. {\bf{13}} (1987) 218-222 (in Russian).


\bibitem {Kh91}   Khaliulilin Kh.F., Khodykin S.A. and Zakharov A.I.

\textit{On the nature of the anomalously slow apsidal motion  of DI
Herculis}

Astrophys.J. {\bf{375}} (1991) 314-320.



\bibitem {KhKh92}   Khaliulilina A.I. and Khaliulilin Kh.F.

\textit{Vrashenie linii apsid v zatmennoi dvoinoi sisteme  HS Her}

Astron.cirk.N 1552 (1992) 15-16(in Russian).


\bibitem {Mart88}   Martynov D.Ia., Voloshina I.E. and Khaliulilina A.I.

\textit{Fotometricheskie elementy zatmennoi sistemy  HS Her}

Asrton. zh. {\bf{65}} (1988) 1225-1229 (in Russian).



\bibitem {Mezz}   Mezzetti M., Predolin F., Giuricin G. and Mardirossian F.

\textit{Revised photometric elements of eight eclipsing binaries}

Astron.Astroph.Suppl. {\bf{42}} (1980) 15-22.


\bibitem {Mossak87a}   Mossakovskaya L.V. and Khaliulilin Kh.F.

\textit{Tret'e telo v zatmennoi sisteme s apsidalnym dvizheniem CO
Lac?}

Astron. cirk.N1495, 5-6 (in Russian).


\bibitem {Sem}   Semeniuk I.

\textit{Apsidal motion in binary systems. I. CO Lacertae, an
eclipsing variable with apsidal motion}

Acta Astron. {\bf{17}} (1967) 223-224.



\bibitem {Anders91}   Andersen J.

\textit{Accurate masses and radii of normal stars}

Astron.Astroph.Rev. {\bf{3}} (1991) 91-126.



\bibitem {Kh85}   Khaliulilina A.I., Khaliulilin Kh.F. and Martynov D.Ya.

\textit{Apsidal motion and the third body in the system RU
Monocerotis}

Montly .Not.Roy.Astron.Soc. {\bf{216}} (1985) 909-922.




\bibitem {Mart86}   Martynov D.Ya. and Khaliulilina A.I.

\textit{RU Monocerotis: poslednie resultaty}

Astron.zh.{\bf{63}} (1986) 288-297 (in Russian).






\bibitem {Sh62}   Shneller H.

\textit{Uber die periodenanderrungen bei bedeckungsveranderlichen}

Budd.Mitt. N53 (1962) 3-41.


\bibitem {Kort}   Kort S.J., J. de,
\textit{The orbit and motion of priastron of GN Normae}

Ricerche Astron. {\bf{3}} (1954) 119-128.



\bibitem {Kamp}   Kamper B.C.

\textit{Light-time orbit and apsidal motion of eclipsing binary U
Ophiuchi}

Astrophys. Space Sci. {\bf{120}} (1986) 167-189.


\bibitem {Claus86}   Clausen J.V., Gemenez A. and Scarfe C.

\textit{Absolute dimentions of eclipsing binaries.XI. V 451
Ophiuchi}

Astron.Astroph. {\bf{167}} (1986) 287-296.





\bibitem {KhKo89}   Khaliulilin Kh.F. and Kozyreva V.S.

\textit{Photometric light  curves and physical parameters of
eclipsing binary systems IT Cas, CO Cep, AI Hya with possible
apsidal motions}

Astrophys. and Space Sci., {\bf{155}}  (1989) 53-69.




\bibitem {Monet}   Monet D.G.

\textit{A discussion of apsidal motion detected in selected
spectroscopic  binary systems}

Astrophys. J., {\bf{237}}  (1980) 513-528.



\bibitem {Sv}   Svechnicov M.A.

\textit{Katalog orbitalnyh elementov, mass i svetimostei tesnyh
dvoinyh zvezd}

Irkutsk, Izd-vo Irkutsk. Univer.(In Russian).




\bibitem {Br}   Brancewicz H.K. and Dworak T.Z.

\textit{A Catalogue of parameters for eclipsing binaries}

Acta Astron., {\bf{30}}  (1980) 501-524.




\bibitem {Wolf95}   Wolf M. and Saronova L.,

\textit{Aspidal motion in the eclipsing binary FT Ori}

Astron.Astroph. Suppl.{\bf{114}} (1995) 143-146.



\bibitem {Dr}   Drozdz M., Krzesinski J. and Paydosz G.,

\textit{Aspidal motion of IQ Persei}

Inf. Bull.Var.Stars, N3494, 1-4.


\bibitem {Lacy}   Lacy C.H.S. and Fruch M.L.

\textit{Absolute dimentions and masses of eclipsing binaries. V. IQ
Persei}

Astroph.J.{\bf{295}} (1985) 569-579.


\bibitem {Anders83}   Andersen J.

\textit{Spectroscopic observations of eclipsing binaries.V. Accurate
mass determination for the B-type systems V 539 Arae and $\xi$
Phaenicis}

Astron.Astroph.{\bf{118}} (1983) 255-261.


\bibitem {Odell}   Odell A.P.

\textit{The structure of Alpha Virginis.II. The apsidal constant}

Astroph.J.{\bf{192}} (1974) 417-424.




\bibitem {Gron}   Gronbech B.

\textit{Four-color photometry of eclipsing binaries.V. photometric
elements of NO Puppis }

Astron.Astroph.{\bf{50}} (1980) 79-84.




\bibitem {Harm}   Harmanec P.

\textit{Stellar masses and radii based on motion binary data}

Bull.Astron.Inst.Czech.{\bf{39}} (1988) 329-345.



\bibitem {Anders84}   Andersen J., Clausen L.V. and Nordstrom B.

\textit{Absolute dinemtions of eclipsing binaries.V. VV Pyxidis a
detached early A-tipe system with equal components}

Astron.Astroph.{\bf{134}} (1984) 147-157.


\bibitem {lacy93b}   Lacy C.H.S.

\textit{The photometric orbit and apsidal motion of YY Sagittarii}

Astroph.J.{\bf{105}} (1993) 637-645.



\bibitem {lacy93a}   Lacy C.H.S.

\textit{The photometric orbit and apsidal motion of V 523
Sagittarii}

Astroph.J.{\bf{105}} (1993) 630-636.



\bibitem {lacy93c}   Lacy C.H.S.

\textit{The photometric orbit and apsidal motion of V 526
Sagittarii}

Astroph.J.{\bf{105}} (1993) 1096-1102.



\bibitem {Anders85}   Andersen J. and Gimenes A.

\textit{Absolute dinemtions of eclipsing binaries.VII. V 1647
Sagittarii}

Astron.Astroph.{\bf{145}} (1985) 206-214.



\bibitem {Sw}   Swope H.H.

\textit{V 2283 Sgr, an  eclipsing star with rotating apse}

Ric.Astron.{\bf{8}} (1974) 481-490.




\bibitem {O'K}   O'Konnell D.J.K.

\textit{The photometric orbit and apsidal motion of V2283
Sagittarii}

Ric.Astron.{\bf{8}} (1974) 491-497.




\bibitem {Anders85b}   Andersen J., Clausen L.V., Nordstrom B. and
Popper D.M.

\textit{Absolute dinemtions of eclipsing binaries.VIII. V 760
Scorpii}

Astron.Astroph.{\bf{151}} (1985) 329-339.




\bibitem {Claus95}   Clausen L.V., Gimenez A.  and
Houten C.J.

\textit{Four-color photometry of eclipsing binaries.XXVII. A
photometric anallysis of the (possible ) Ap system AO Velorum}

Astron.Astroph.{\bf{302}} (1995) 79-84.


\bibitem {Popp80}   Popper D.M.

\textit{Stellar masses}

Ann. Rev. Astron. and Astroph.{\bf{18}} (1980) 115-164.



\bibitem {Dukes}   Dukesr R.J.

\textit{The beta Cephei nature of Spica}

Astroph.J.{\bf{192}} (1974) 81-91.




\bibitem {Kh87}   Khaliulilina A.I.

\textit{DR Vulpeculae: the quadruple system}

Montly .Not.Roy.Astron.Soc. {\bf{225}} (1987) 425-436.




\bibitem {KhKh88}
Khaliulilin Kh.F. and Khaliulilina A.I.

\textit{Fotometricheskoe issledovanie zatmennoi zvezdy DR Vul.
Parametry sistemy i vrashenie linii apsid},

Astron.zh., {\bf N{65}}  (1988) 108-116 (in Russian).


\bibitem {KhKo89}   Khaliulilin Kh.F. and Kozyreva V.S.

\textit{Photometric light curves and physical parameters of
eclipsing binary systems IT Cas, CO Cep, AI Hya with possible
apsidal motions}

Astrophys. and Space Sci., {\bf{155}}  (1989) 53-69.


\bibitem {Holm}   Holmgren D. anf Wolf M.
\textit{Apsidal motion of the eclipsing binary  IT Cassiopeiae}

Observatory {\bf{116}}  (1996) 307-312.


\end{thebibliography}
\end{document}